\documentclass[12pt]{iopart}
\usepackage{iopams}  
\usepackage{graphicx,epsfig}
\usepackage{hyperref}
\newcommand{\beq}{\begin{equation}}
\newcommand{\eeq}{\end{equation}}
\newcommand{\vx}{\mathbf{x}}

\newcommand{\ei}{\end{itemize}}
\newcommand{\vk}{\mathbf{k}}
\newcommand{\vko}{\mathbf{k}_1}
\newcommand{\vkt}{\mathbf{k}_2}

\newcommand{\vkth}{\mathbf{k}_3}
\newcommand{\vkf}{\mathbf{k}_4}

\newcommand{\vq}{\mathbf{q}}

\begin{document}
\title{Kurtosis of height fluctuations in $(1+1)$ dimensional KPZ Dynamics}
\author{Tapas Singha \& Malay K. Nandy}
\address{Department of Physics, Indian Institute of Technology Guwahati, Guwahati 781039, India.}
\ead{s.tapas@iitg.ernet.in \& mknandy@iitg.ernet.in}
\vspace{10pt}
%%%%%%%%%%%%%%%%%%%%%%%%%%%%%%%%%%%%%%%%%%%%%%%%%%%%%%%%%%%%%%%%%%%%%%%%%%%%%%%%%%%%%%%%%%%%%%%%%

% \begin{document}
% \title{Kurtosis of height fluctuations in $(1+1)$ dimensional KPZ Dynamics}
% \author{Tapas Singha} 
% \email{s.tapas@iitg.ernet.in}
% \author{Malay K. Nandy}
% \email{mknandy@iitg.ernet.in}
% \affiliation{Department of Physics, Indian Institute of Technology
%   Guwahati, Guwahati 781039,  India.}
% \date{\today}

\begin{abstract}
We study the fourth order normalized cumulant of height fluctuations governed by $1+1$ dimensional Kardar-Parisi-Zhang (KPZ) equation 
for a growing surface. Following a diagrammatic renormalization scheme, we evaluate the kurtosis $Q$ from the connected diagrams leading
to the value $Q=0.1523$ in the large-scale long-time limit.
\end{abstract}
% %\pacs{81.15.Aa, 68.35.Fx, 64.60.Ht, 05.10.Cc}
% % \maketitle
% %Keywords: Kinetic roughening (Theory), Self-affine roughness (Theory), Dynamical processes (Theory), Stochastic processes (Theory).
\hspace{4pc}
\noindent{\bf Keywords}: Kinetic roughening (Theory), Self-affine roughness (Theory), \hspace{4pc} Dynamical processes (Theory), Stochastic processes (Theory).

\section{Introduction}
Growth of a surface or interface has been one of the most important and
well-studied fields in nonequilibrium statistical physics since a long time \cite{book_stanley,krug97,Halpin95,Meakin93,Family_Physica}. Kardar,
Parisi and Zhang \cite{KPZ86} first proposed a paradigmatic nonlinear equation for local surface growth capable of describing many growth phenomena.
The equation, called the KPZ equation, is
expressed as 
\beq
\frac{\partial h} {\partial t}=\nu_0
\nabla^{2} h+\frac{\lambda_0}{2} (\nabla h)^{2}+ \eta,
\label{eq-kpz}
\eeq
where $h(\vx,t)$ is the field of height fluctuations, $\nu_0$ is the surface
tension that relaxes particles from local maxima to local minima, and $\lambda_0$
is the strength of local interaction. Here $\eta(\vx,t)$ is the
deposition noise with zero average, $\langle \eta(\vx,t) \rangle=0$,
and its covariance is modeled as a short range correlation
\beq
\langle \eta(\vx,t) \, \eta(\vx',t')\rangle=2 D_0 \, \delta^d(\vx-\vx') \, \delta(t-t'),
\eeq
with $d$ the substrate dimension. 
 
The roughness of a self-affine surface is characterized by the  width  $w$ of the
interface (or standard deviation $w$ of the height fluctuations), given by the dynamic scaling relation
\beq
w(L,t) \sim L^{\chi} f\left(\frac{t}{L^z}\right),
\eeq
as suggested by Family and Vicsek \cite{Family_Physica},
where $L$ is the size of the interface and $f(\cdot)$ is a universal function having asymptotics such that
$ w(L,t) \sim  t^{\beta} $ when  $t \ll L^z$ and $ w(L,t) \sim  L^{\chi} $ for $ t\gg L^z$.
Here the exponent $\chi$ characterizes the roughness of the surface, $z$ is the dynamic exponent, and the ratio $\beta=\frac{\chi}{z}$ is known 
as the growth exponent.  The roughness exponent $\chi$ is an important parameter in experiments; adsorption, catalysis \cite{Pfeifer_83} and
optical properties \cite{Moskovits_85} of a thin film are affected by the roughness of the surface. The exponents are related via the 
scaling relation $\chi+z=2$ 
\cite{Meakin86,Krug_PRA.36.5465,KPZ89}, which is independent of the substrate
dimension. 
 
 There are various growth phenomena that are
believed to be in the KPZ universality class on the basis of the numerical
values of the scaling exponents \cite{book_stanley}. A few of them are thin
film deposition \cite{PaivaReis_07}, bacteria colony
growth \cite{Vicsek_90,Huergo_10}, fluid flow in porous media \cite{Rubio_89},
turbulent liquid crystal \cite{K_A_Takeuchi,kazumasa}, one dimensional
polynuclear growth (PNG) \cite{saarloos_86,goldenfeld,krug_pra_88,Michael_Herbert_00},
slow combustion of a sheet of paper
\cite{Myllys_PRE.64.036101,miettinena_epjb_46_55_05}. In addition, many problems
are equivalent to the KPZ equation, e.g., Burgers
equation \cite{forster_pra_16_732_77} describes the vorticity free velocity,
directed polymer in random 
media \cite{kpz87,Fisher_PRB_43_10728} and in random potentials (DPRP)
\cite{J-krug_92}, sequence alignment of gene or 
protein \cite{Hwa_Lassig_PRL_76_2591,Hwa_gene_allignment_nature_399}, heat
equation of multiplicative noise obtained via the Cole-Hopf
transformation of the KPZ equation. 

A renormalization group (RG) analysis \cite{KPZ86}
of the $1+1$ dimensional KPZ equation yielded roughness exponent
$\chi=\frac{1}{2}$ and dynamic exponent
$z=\frac{3}{2}$ which are consistent with various numerical models, e.g.,
ballistic deposition \cite{Meakin86,Family_Physica}, Eden model
\cite{M_Eden,plischke_prl_84,jullien_85,eden_plischke_85}, restricted solid on
solid model (RSOS) \cite{Meakin93}, single step model (SSM) \cite{Meakin86,Plischek_87}. RG
calculation in $2+1$ dimensions is unable to yield the exponents of KPZ
equation. A number of analytical techniques have been employed to
study the
higher dimensional scaling exponents of KPZ equation, e.g., the mode coupling
 \cite{Colaiori_PRL_86_3946,Beijeren_PRL_54_2026},
the operator product expansion \cite{Lassig_PRL_80_2366}, the self-consistent
expansion \cite{Schwartz_Edwards_1992} and the non-perturbative RG
\cite{Canet_PRL_104_150601,Kloss_PRE2012}.

Although many physical problems are described by the 
$1+1$ dimensional KPZ equation,  enough attention has not been payed to understand the statistical 
probability distribution of the height fluctuations as it is a challenging task to obtain it directly 
from the KPZ equation. Nevertheless, it is a highly desirable objective and it demands extensive 
theoretical and experimental studies \cite{Meakin93}.
Pr{\"a}hofer and Spohn \cite{Michael_Herbert_00} studied the PNG model for three different initial 
conditions namely flat, droplet
and stationary self similar which are in the KPZ universality class on the basis
of scaling exponents. For the droplet initial condition the distribution was
found to be the Gaussian unitary ensemble (GUE) Tracy-Widom (TW) whereas for the
flat initial condition it is the Gaussian orthogonal 
ensemble (GOE) Tracy-Widom (TW) distribution. The height distribution was
modeled through the relation $h(x, t)\sim v_{\infty} t
+ (\Gamma t)^{1/3} \phi $ (with the parameter
$\Gamma=\frac{D_0^2 \lambda_0}{8 \nu_0^2}$) where $\phi $ is a random variable
determined by appropriate random matrices and $v_{\infty}$ is the rate of growth
at long times \cite{K_A_Takeuchi}.

Imamura and Sasamoto \cite{Imamura_prl_108_2012} considered a Brownian motion as
an initial condition from both sides of the substrate and used the Bethe ansatz and a replica trick
to find the height distribution in terms of a Fredholm determinant. Takeuchi \cite{Takeuchi_prl_110_2013}
obtained scaling functions smoothly connecting the crossover between the GOE-TW and Baik-Rains $F_0$ distributions. 
He observed that the moments pass through minimum values, known as the Takeuchi minima. Such behavior was also noticed 
via TLC experimental studies apart from the numerical study of the PNG model.

In spite of the same scaling exponents, the statistical behavior of stochastic
processes may be different due to difference in the underlying probability
distribution functions (pdf) that incorporate the basic features of a
dynamical process \cite{PhysRevE.64.036110}. In principle, the pdf can be
calculated analytically by solving the Fokker-Planck
equation \cite{huse_85,Parisi_1990} corresponding to the KPZ equation. However, this is practically infeasible due to the nonlinear term.
As an alternative, a few
higher order moments can be calculated to understand the statistical behavior of
the interface and the corresponding universality class. It may be noted that the
 measurement accuracy of
higher order moments is greater than that of the scaling exponents
\cite{Reis_pre_70_031603_05} in experiments.

In this paper, we consider the $1+1$ dimensional KPZ equation with a flat
substrate  and focus on the evaluation of the fourth order cumulant
relevant to the kurtosis in stationary state.  We apply a diagrammatic
approach to find a renormalized expression for the loop integral
corresponding to the fourth order cumulant in the large scale long
time limit.

The paper is organized as follows. Section 2 defines the moments and cumulants and states the relations between them.
 Section 3 is devoted to the calculation of fourth order cumulant.  In Section 4, the calculation of excess kurtosis is presented.
 Section 5 presents a discussion and conclusion and a comparison the kurtosis value with other numerical and experimental findings.

\section{Moments and Cumulants}

Higher order moments and cumulants are usually defined in terms of a generating function.  The moment generating function $Z(\beta)$ is defined as
\beq
Z(\beta)\equiv \langle e^{\beta h} \rangle=\sum^{\infty}_{n=0}
\frac{\beta^n}{n!} \langle h^n\rangle,
\eeq
where $\langle h^n\rangle$ is the $n$th order moment of a random
variable $h$, given by 
\beq
\langle h^n \rangle =\int^{+\infty}_{-\infty} h^n P(h)\,dh,
\eeq
for a given normalized probability distribution function $P(h)$.
The cumulant generating function $F(\beta)$ is defined as 
\beq
F(\beta)=\ln Z(\beta) =\sum^{\infty}_{n=1}
\frac{\beta^n}{n!} \langle h^n\rangle_c,
\eeq
where $\langle h^n \rangle_c$ is the $n$th order cumulant. 
The moments are related to the cumulants as
\begin{eqnarray}
\langle h \rangle&=&\langle h \rangle_c  \nonumber \\
\langle h^2 \rangle&=&\langle h^2 \rangle_c + \langle h \rangle^2_c \nonumber
\\
\langle h^3 \rangle&=&\langle h^3 \rangle_c +3 \langle h \rangle_c\langle h^2
\rangle_c +\langle h \rangle^3_c \nonumber \\
\langle h^4 \rangle&=&\langle h^4 \rangle_c +4\langle h \rangle_c \langle h^3
\rangle_c+3 \langle h^2 \rangle^2_c +6\langle h \rangle^2_c \langle h^2
\rangle_c + \langle h \rangle^4_c.
\end{eqnarray}
 The fluctuating interface field $h(\vx,t)$ has a zero mean;
$\langle h(\vx,t) \rangle=\langle h(\vx,t) \rangle_c=0$. Hence the second and fourth order cumulants \cite{J-krug_92,PhysRevE.64.036110} 
are related to the
moments as
\beq
\langle h^2 \rangle_c=\langle h^2 \rangle,
\eeq
and
\beq
\langle h^4 \rangle_c= \langle h^4 \rangle-3 \langle h^2 \rangle^2.
\eeq
Kurtosis $Q$ is defined by the ``normalized'' fourth order cumulant as
\beq
Q=\frac{\langle h^4
\rangle_c}{\langle h^2 \rangle^2_c}=\frac{\langle h^4 \rangle}{\langle h^2
\rangle^2}-3.
\label{defh4}
\eeq
In the diagrammatic approach, a connected diagram with $n$ external legs
corresponds to the $n$th order cumulant. Henceforth, we focus on the fourth cumulant instead of the moment 
and therefore we evaluate the connected loop diagram to obtain the cumulant in order to evaluate the kurtosis.

\section{The Fourth-order Cumulant}
The fourth order cumulant $\langle h^4(\vx,t)\rangle_c$ of the height
fluctuations measures the degree of flatness of the probability
distribution.  Using the Fourier transform 
\beq
h(\vx,t)=\int \frac{d^{d}k \, d\omega}{(2\pi)^{d+1}} \, h(\vk,\omega) \,
e^{i(\vk \cdot \vx-\omega t)},
\eeq
the KPZ equation (\ref{eq-kpz}) is written in momentum and frequency space as
\begin{equation}
(-i \omega+\nu_0 k^2) \, h(\vk,\omega)=\eta(\vk,\omega)
-\frac{\lambda_0}{2} \int\!\! \frac{d^dq \, d\Omega}{(2\pi)^{d+1}}
\,[\vq\cdot(\vk-\vq)]\,h(\vq,\Omega)\,h(\vk-\vq,\omega-\Omega).
\label{KPZFT}
\end{equation}
This equation forms the basis of developing a perturbation theory.

\begin{figure}[h!t]
\begin{center}
\includegraphics[width=4.3cm]{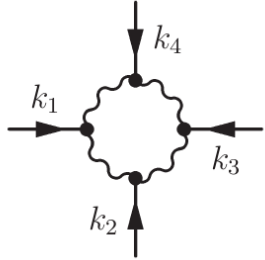}
\end{center}
% %\vskip1cm
\caption{Feynman Diagram corresponding to the fourth cumulant}
\end{figure}

The expression for the fourth cumulant is given by the connected fourth order
correlation in momentum and frequency space as
\begin{eqnarray}
\langle h^4(\vx,t) \rangle_c &=& \int \frac{d^{d+1} \hat{k}_1}{(2\pi)^{d+1}}\int
\frac{d^{d+1}\hat{k}_2}{(2\pi)^{d+1}} \!\int
\frac{d^{d+1} \hat{k}_3}{(2\pi)^{d+1}}
\!\int \frac{d^{d+1}\hat{k}_4}{(2\pi)^{d+1}}
\nonumber \\
&&
\langle h(\hat{k}_1) h(\hat{k}_2) h(\hat{k}_3) h(\hat{k}_4)\rangle_c 
\,e^{i(\hat{k}_1+\hat{k}_2+\hat{k}_3+\hat{k}_4)\cdot \hat{x}},
\label{h4ini}
\end{eqnarray}
where $\hat{x}\equiv (\vx,t)$ and $\hat{k}_i\equiv (\vk_i,\omega_i)$ are vectors in $(d+1)$-vector notation.
We use the diagrammatic approach and obtain a connected loop diagram for the
fourth cumulant as shown in Fig.~1, where a wiggly line represents the
correlation and a solid line the  response function.
We express the correlation in accordance with the connected loop diagram in 
Fig.~1, so that
\begin{eqnarray}
\langle h^4(\vx,t)\rangle_c &=& \int\frac{d^{d+1}\hat{k}_1}{(2\pi)^{d+1}}\int
\frac{d^{d+1} \hat{k}_2}{(2\pi)^{d+1}} \!\int
\frac{d^{d+1} \hat{k}_3}{(2\pi)^{d+1}} \,G(\hat{k}_1)\,G(\hat{k}_2)\, G(\hat{k}_3)
\nonumber\\
&& L_4(\hat{k}_1,\hat{k}_2,\hat{k}_3) \, G(-\hat{k}_1-\hat{k}_2-\hat{k}_3),
\label{h4L4}
\end{eqnarray}
where $L_4$ represents  the renormalized loop corresponding to the amputated part of the diagram (excluding the external legs).
Its bare value is written as 
\begin{eqnarray}
L^{(0)}_4(\hat{k}_1,\hat{k}_2,\hat{k}_3) &=&
16 \left(-\frac{\lambda_0}{2}\right)^4 (2D_0)^4
\int \frac{d^{d+1}\hat{q}}{(2\pi)^{d+1}} [\vq
\cdot(\vq-\vko)][\vq\cdot(\vkt+\vq)]
\nonumber\\
&&
[(\vq+\vkt)\cdot(\vkth+\vkt+\vq)]
[(\vkf+\vkth+\vkt+\vq)\cdot(\vq+\vkt+\vkth)] \nonumber \\
&& G_0(\hat{q})G_0(-\hat{q}+\hat{k}_1) G_0(-\hat{q}) G_0(\hat{q}+\hat
{k}_2) 
G_0(-\hat{q}-\hat{k}_2) G_0(\hat{q}+\hat{k}_3+\hat{k}_2)
\nonumber \\
&&
G_0(\hat{q}+\hat{k}_2+\hat{k}_3+\hat{k}_4)G_0(-\hat{q}-\hat{k}_2-\hat{k}_3),
\label{L4}
\end{eqnarray}
where the prefactor 16 is a combinatorial factor.

Now we evaluate Eq.\ \ref{L4} by
performing frequency and momentum integrations where the momentum is
restricted to the shell
$\Lambda_0 e^{-r} \leq q\leq \Lambda_0$. This leads to   
\beq
L^<_4(r)=\frac{5}{2} K_d \frac{\lambda_0^4 D^4_0 }{\nu_0^{7} \Lambda_0^{5}}\left
(\frac{e^{5r}-1}{5}\right),
\label{L4r}
\eeq
where $K_d=\frac{S_d}{(2\pi)^d}$, with $S_d$ the surface area of a unit sphere in $d$ dimensional space.

We follow Yakhot and Orszag's \cite{yakhot_prl_57_1772_86,yakhot_j_s_comput_1_3_86} scheme
of renormalization without rescaling and obtain from Eq.\ \ref{L4r} the
differential equation
\beq
\frac{dL_4}{dr}=\frac{5}{2\pi} \frac{\lambda_0^4 D^4(r)}{ \nu^7(r)\Lambda^{5}(r)},
\label{dL4}
\eeq
in one dimension.

Noting that $\Lambda(r)=\Lambda_0e^{-r}$ and using the asymptotic expressions 
\beq
\nu(r)=\lambda_0 \sqrt{\frac{D_0}{2\pi \nu_0 \Lambda_0}} \, e^{r/2}
\label{nur}
\eeq
and
\beq
D(r)=\frac{\lambda_0 D_0}{\nu_0} \sqrt{\frac{D_0}{2\pi \nu_0 \Lambda_0}} \, e^{r/2}
\label{Dr}
\eeq
obtained from renormalization-group calculations \cite{skew2014},
we obtain the solution to the differential equation (\ref{dL4}) as
\beq
L_4(r)= \frac{10\pi}{7} \lambda_0 \left(\frac{2 D^5_0}{\pi \nu^5_0
\Lambda^7_0}\right)^{1/2} e^{7r/2}.
\eeq
The factor $\Lambda_0 e^{-r}$ is interpreted as a momentum $k_i$ in the large
scale limit. Since the loop integral must be symmetric with respect
to interchange of the external momenta $k_1$, $k_2$ and $k_3$, we
replace $\Lambda_0 e^{-r}$ by the fully symmetric form $k^{1/3}_1
k^{1/3}_2 k^{1/3}_3$. Thus we have 
\beq
L_4(\vk_1,0;\vk_2,0;\vk_3,0)=\frac{10 \pi}{7} \lambda_0 \left(\frac{2
D^5_0}{\pi \nu^5_0}\right)^{1/2} k^{-7/6}_1 k^{-7/6}_2 k^{-7/6}_3,
\label{remainingint}
\eeq
when the external frequencies are zero.
To obtain the frequency dependency, we take the scaling 
function as
\beq
k_{i}^{7/6}f_{4}\left(\frac{\omega_i}{k_{i}^z}\right) =\frac{1}{k_{i}^{17/6}
\nu^2(k_{i}) |G(\vk_{i},\omega_{i})|^2},
\label{dynamic}
\eeq
where the renormalized response function is given
by $G(\vk_i,\omega_i)=[-i \omega_i+\nu(k_i)\vk_i^2]^{-1}$ with the renormalized surface tension 
\beq
\nu(k_i)=\lambda_0 \sqrt{\frac{D_0}{2\pi \nu_0}} \, k_i^{-1/2}.
\label{nufk}
\eeq
The above scaling relation (\ref{dynamic}) obeys consistency with the zero frequency limit and with 
the real valuedness of $\Lambda_0 e^{-r}$, in addition to being of the correct dimension.
Now with the aid of Eq.~(\ref{dynamic}),
Eq.~(\ref{remainingint}) is modified to the frequency dependent form 
\begin{eqnarray}
L_4(\hat{k}_1,\hat{k}_2,\hat{k}_3) &=& \frac{10 \pi}{7} \lambda_0 \left(\frac{2
D^5_0}{\pi \nu^5_0}\right)^{1/2} k_1^{17/6} k_2^{17/6} k_3^{17/6} \,
\nu^2(k_1) |G(k_1,\omega_1)|^2 \nonumber \\
&&\nu^2(k_2) |G(k_2,\omega_2)|^2\,\nu^2(k_3)|G(k_3,\omega_3)|^2,
\label{L4renorma}
\end{eqnarray}
representing the renormalized loop diagram in Fig.~1. We substitute the
renormalized expression for the loop diagram from Eq.~\ref{L4renorma} in Eq.\
\ref{h4L4} and treat the external legs representing  the response functions as
renormalized. Carring out the frequency integration,
we obtain  
\begin{equation}
\langle h^4(\vx,t)\rangle_c=\frac{10}{7}
\left(\frac{D_0}{2\pi \nu_0}\right)^2 \int^{\infty}_{-\infty} 
dk_1 \int^{\infty}_{-\infty}
d k_2 \!\int^{\infty}_{-\infty} d k_3 \, \Phi(k_1,k_2,k_3)
\label{h4I}
\end{equation}
where 
\begin{equation}
\Phi(k_1,k_2,k_3)=\frac{u(k_1,k_2,k_3)}{v(k_1,k_2,k_3)}
\end{equation}
with
\begin{eqnarray}
u(k_1, k_2, k_3)&=&
(3(|\vko|^{9/2}+|\vkt|^{9/2}+|\vkth|^{9/2})+5(|\vko|^{3/2}+|\vkt|^{3/2}+|\vkth|^
{3/2})|\vko+\vkt+\vkth|^3 \nonumber \\ &&
+7(|\vko|^3+|\vkt|^3+|\vkth|^3)|\vko+\vkt+\vkth|^{3/2}+17(|\vko|^3|\vkt|^{3/2} +
|\vko|^{3/2}|\vkt|^{3} \nonumber \\
&& +|\vko|^3|\vkth|^{3/2}+|\vko|^{3/2}|\vkth|^{3}+|\vkt|^{3}
|\vkth|^{3/2}+|\vkt|^{3/2} |\vkth|^{3}) \nonumber \\
&& +22(|\vko|^{3/2}|\vkt|^{3/2}+|\vko|^{3/2}|\vkth|^{3/2}+|\vkt|^{3/2}
|\vkth|^{3/2})|\vko+\vkt+\vkth|^{3/2} \nonumber \\
&& +90|\vko|^{3/2}|\vkt|^{3/2}|\vkth|^{3/2}+|\vko+\vkt+\vkth|^{9/2})
\end{eqnarray}
and
\beq
v(k_1, k_2, k_3)= 64 |\vko|^{7/6} |\vkt|^{7/6} |\vkth|^{7/6} 
(|\vko|^{3/2}+|\vkt|^{3/2}+|\vkth|^{3/2}+|\vko + \vkt + \vkth|^{3/2})^4.
\eeq
The symmetry of the function  $\Phi(k_1,k_2,k_3)$ allows us to write Eq.~\ref{h4I} as
\begin{equation}
\langle h^4(\vx,t)\rangle_c= \frac{10}{7} \left(\frac{D_0}{2\pi
\nu_0}\right)^2
\int^{\infty}_{\mu} dk_1  \int^{\infty}_{\mu}
d k_2 \!\int^{\infty}_{\mu} dk_3 
\left[2 \Phi(k_1,k_2,k_3)+ 6 \Phi(-k_1,k_2,k_3)\right]
\label{h4spI}
\end{equation}
where $\mu$ is an infrared cut off. Now we write the integrations separately as
\beq
I_1(\mu)= \int^{\infty}_{\mu}  dk_1 \int^{\infty}_{\mu}
d k_2 \!\int^{\infty}_{\mu} d k_3 \,\Phi(k_1,k_2,k_3)
\label{I1mu}
\eeq
and 
\beq
I_2(\mu)= \int^{\infty}_{\mu}  dk_1 \int^{\infty}_{\mu}
d k_2 \!\int^{\infty}_{\mu} d k_3 \,\Phi(-k_1,k_2,k_3).
\label{I2mu}
\eeq
These integrals are expected to be of the forms
\begin{eqnarray}
I_1(\mu)= a_1 \mu^{-2} \label{eq:I1}\\
I_2(\mu)=a_2 \mu^{-2}. \label{eq:I2}
\end{eqnarray}
where $a_1$ and $a_2$ are dimensionless constants.
Substituting Eqs.~\ref{eq:I1} and \ref{eq:I2} in Eq.~\ref{h4spI} yields
\beq
\langle h^4(\vx,t)\rangle_c=\frac{10}{7}
[2 a_1+6 a_2] \left(\frac{D_0}{2\pi \nu_0}\right)^2  \frac{1}{\mu^2}
\label{h4M1M2}
\eeq
We evaluate  the integrals in Eqs.~\ref{I1mu} and \ref{I2mu} by integrating over
$k_1$, $k_2$ and $k_3$ numerically and obtain the dimensionless constants as
\beq
a_1= \lim_{\mu \rightarrow 0^+}[\mu^2 I_1(\mu)]=0.007505
\label{a_1}
\eeq
and
\beq
a_2= \lim_{\mu \rightarrow 0^+}[\mu^2 I_2(\mu)]=0.026297,
\label{a_2}
\eeq
the numerical values having converged to the above values for decreasing values of the parameter $\mu$ very close to zero.

\section{The Kurtosis}

Having calculated the fourth-order cumulant given by Eq.~(\ref{h4M1M2}), 
we now need the value of the second-order cumulant to determine the value of kurtosis given by Eq.~(\ref{defh4}).
The second moment is written as 
\beq
\langle h^2(\vx,t)\rangle =\int \frac{d^dk_1 \, d
\omega_1}{(2\pi)^{d+1}} \int
\frac{d^dk_2\, d\omega_2}{(2\pi)^{d+1}} \,\langle
 h(\vk_1,\omega_1)h(\vk_2,\omega_2)\rangle\,
e^{i(\vk_1+\vk_2) \cdot \vx} \,e^{-i(\omega_1+\omega_2)t}.
\label{h^2}
\eeq

\begin{figure}[h!t]
\begin{center}
\includegraphics[width=4.6cm]{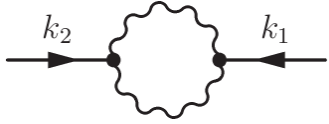}
\end{center}
%\vskip1cm
\caption{Feynman diagram corresponding to the second moment}
\end{figure}

\noindent
Homogeneity in space and time implies that the correlation takes the form
\beq
\langle h(\vk_1,\omega_1) \, h(\vk_2,\omega_2)\rangle = Q(\vk_1,\omega_1)\,
(2\pi)^d\delta^d(\vk_1+\vk_2)\, (2\pi)\delta(\omega_1+\omega_2).
\eeq
The correlation function $Q(\vk,\omega)$ can be written in accordance with the one-loop diagram in Fig.~2, so that
\beq
\langle h^2(\vx,t)\rangle_c =
\langle h^2(\vx,t)\rangle = \int \frac{d^dk \, d\omega}{(2\pi)^{d+1}}
\,G(\vk,\omega) \, L_2(\vk,\omega) \, G(-\vk,-\omega)
\label{W2FT}
\eeq
The renormalized value of $L_2$, given by the loop integral coming from the amputated part of Fig.~2, can be obtained as
\beq
L_2(\vk,\omega)= \frac{\lambda^2_0 D^2_0}{\pi \nu^2_0} \, k \, |G(\vk,\omega)|,
\label{L2_k}
\eeq
by means of a renormalization scheme as shown in Ref.~\cite{skew2014}.  This yields the second cumulant as
\beq
\langle h^2(\vx,t) \rangle_c=\frac{4}{\pi} \left(\frac{D_0}{2 \pi \nu_0}\right)
\frac{1}{\mu}.
\label{eq=Q2}
\eeq

Employing Eqs.~(\ref{h4I}) and (\ref{eq=Q2}) in the definition of kurtosis leads to
\beq
Q=\frac{\langle h^4(\vx,t)\rangle_c}{\langle
h^2(\vx,t)\rangle^2_c}=\frac{5\pi^2}{28}(a_1+3a_2),
\eeq
free from the infra-red cutoff $\mu$ and the model parameters $\nu_0$, $D_0$, and $\lambda_0$.
Employing the numerical values of $a_1$ and $a_2$ from Eqs.~\ref{a_1} and \ref{a_2} yields the kurtosis as
\beq
Q=0.152267.
\label{Q_1517}
\eeq
We compare this value with other stationary values of kurtosis in Table~1.

\begin{table}[ht]
\caption{Values of Kurtosis in $(1+1)$ dimensions.}
\begin{center}
\begin{tabular}{l|c|c|c}
\hline\hline
{\it System of study} & {\it Method} & {\it Kurtosis} & {\it Reference}\\ 
\hline
PNG (stationary) & Numerical & $0.289$ & \cite{Michael_Herbert_00} \\
\hline
g1 DPRM (stationary) & Numerical & $0.278$ & \cite{Halpin-Healy14}
\\
\hline
KPZ (present calculation) & Analytical & $0.1523$ &  Eq.~(\ref{Q_1517})\\ 
\hline
\end{tabular}
\end{center}
\label{table:nonlin} % is used to refer this table in the text
\end{table}

\section{Discussion and Conclusion}

In this work, we followed a perturbative renormalization approach to
evaluate the Feynman diagram in Fig.~1 that represents the fourth cumulant, noting that contribution to cumulants are given by the connected loop diagrams.
To calculate the connected diagram, we first renormalized its amputated part
in the limit of internal momenta and frequencies much greater than
the external ones and followed Yakhot and Orszag's iterative renormalization scheme  without rescaling \cite{yakhot_j_s_comput_1_3_86}.  This results in a
scale dependent function for the amputated part of the loop diagram at zero
frequency.  Introducing a frequency dependent scaling function that smoothly joins with the obtained scaling behavior and that preserves 
the property of real valuedness, we evaluated the resulting frequency and
momentum integrations to obtain an expression for the fourth order cumulant depending on the infrared cutoff in the momentum integration. 
The evaluation of the second cumulant is simpler which also depends on the the infrared cutoff as shown in Ref.~\cite{skew2014}. 
The behaviors of both the fourth and second order cumulants are seen to be exactly in accordance with the well-known behavior of the $n$th moment,
namely $W_n \sim L^{n \chi}$, where $L$ is the substrate size (so that $\mu\sim L^{-1}$) and $\chi$ is the roughness exponent. 
 These cumulants yield the value of kurtosis $Q=0.1523$ which is free from the infrared cutoff $\mu$ and model dependent
parameters $D_0$ , $\nu_0$, and $\lambda_0$, due to their exact cancellations.

Incidentally, it may be worth mentioning that the
calculation of cumulant amplitudes have been regarded as an important objective by some researchers.  For example,
Krug et al.\ \cite{J-krug_92} studied the growth of an
interface via simulation of the single step model with flat initial
condition and obtained the estimates for the cumulant amplitudes as
$c_2=0.404\pm0.013$ and $c_4=0.020\pm0.002$, suggesting a kurtosis value $Q=\frac{c_4}{c_2^2}=0.123\pm0.020$. Furthermore, they showed that $c_2$ is higher for the steady state case, namely, $c_2=0.712 \pm 0.003$, due to the additional role of fluctuations in the initial conditions besides those operating during the growth.  Earlier, Hwa and Frey \cite{Hwa91} had obtained $c_2=0.69$ via mode-coupling calculations. Tang \cite{Tang92} 
estimated $c_2=0.725\pm 0.005$ using a Monte-Carlo simulation
with the single-step model, giving an excellent estimate for the Baik-Rains
constant \cite{Michael_Herbert_00,Halpin-Healy13}.  A slightly different result, $c_2=0.71$, was obtained through mode-coupling calculation
by Amar and Family  \cite{Amar-Family92}.

In Table~1, our calculated kurtosis value is compared with other stationary values.
Pr\"ahofer and Spohn  \cite{Michael_Herbert_00} studied the PNG model belonging to the $1+1$ dimensional KPZ universality class to find the effect of initial conditions on the statistical behavior of growing interfaces. They obtained different kurtosis values for different initial conditions, namely, $Q=0.0934$ for curved, $Q=0.1652$ for flat, and $Q=0.289$ for stationary initial conditions. Halpin-Healy and Lin \cite{Halpin-Healy14} studied the DPRM with different configurations  such as point-to-point, point-to-line, and stationary cases, leading to GUE-TW, GOE-TW and Baik-Rains $F_0$ distributions, respectively.  Through $g_1$ DPRM, they estimated the stationary kurtosis value to be $Q=0.278$.

% the Takeuchi minimum kurtosis value to be $Q_{TM}=0.117$.

Apart from the values shown in Table~1, there are a few studies where the full Baik-Rains distribution is obtained
in the context of KPZ stationary state. Takeuchi \cite{Takeuchi_prl_110_2013} obtained, via a numerical simulation 
of the PNG model and an experiment on the TLC, a universal function that undergoes a crossover from a transient 
state to the stationary regime. Miettinen {\it et al.\/} \cite{miettinena_epjb_46_55_05} carried out an experimental 
study of the slow propagation of the combustion front on a sheet of paper to investigate upon the front fluctuation distribution. 
Their experimental data for the transient and stationary states were found to fit well with the GOE-TW and $F_0$ distributions, respectively. 
On the other hand, Halpin-Healy and Lin \cite{Halpin-Healy14}, studied the distributions of deposition models (BD, SSM, and RSOS) that strongly
agree with DPRM/SHE results and they turn out to be the Baik-Rains distribution. These investigations establish that the kurtosis value in the
stationary state must be the same as the universal kurtosis of the Baik-Rains distribution 
(for example, $Q=0.289$ \cite{Michael_Herbert_00} or  $0.278$ \cite{Halpin-Healy14}).

In our calculation based on the RG scale elimination scheme, we obtained the renormalized quantities in the large-scale and long-time limits.  
Thus our methodology inexorably selects the stationary regime.  However, the resulting kurtosis value does not agree well with the stationary 
kurtosis value and it is distinctly lower than the Baik-Rains value, $Q=0.28916$.  In our simplified scheme of calculation, we obtained the fourth cumulant at one-loop order.
While a one-loop scheme for the third cumulant was nearly successful in estimating the stationary skewness value \cite{skew2014}, a higher order
calculation would appear to be more appropriate for the fourth cumulant.  Noting that the perturbation expansion is about a Gaussian state,  
the Gaussian behavior seems to play some role with the increase in order of the cumulant, lowering the estimated value.  This opens the door 
for further investigations into the rather 
unexplored stationary KPZ problem that is expected to illuminate upon its semantic relation with Baik-Rains distribution.

We conclude by recalling that there are many growth processes governed by nonequilibrium dynamics that are believed to be in the KPZ universality class on the basis of scaling 
exponents. However, as shown by the PNG and TLC studies, the statistical behavior, that is the probability distribution, depends on the initial 
conditions. The full probability distribution function of the KPZ height fluctuations has never been studied analytically due to the inherent 
difficulty in the problem. However, relevant information about the probability distribution can be obtained through the study a few higher order 
moments and cumulants. Thus the present theoretical study may be viewed as an initial attempt at the quantification of higher order
statistical property of the $1+1$ dimensional KPZ dynamics. We hope that such analytical scheme would be useful for the study of statistical 
behavior of other important stochastic processes as well.

\subsection*{Acknowledgements}

We thank M.~Pr\"ahofer and H.~Spohn for making their data available online \cite{Prahoferdata}. T.~Singha thanks the MHRD, 
Government of India, for financial support through a scholarship.

\end{document}